\documentclass[sigconf, screen]{acmart} 

\renewcommand\footnotetextcopyrightpermission[1]{} 
\usepackage{fancyhdr}
\author{Mingshi Xu} 
\authornote{Corresponding author}
\orcid{0009-0002-2120-4806}
\email{mxuax@connect.ust.hk}
\affiliation{
  \institution{Hong Kong University of Science and Technology}
  \country{Hong Kong SAR}
}

\author{Haoren Zhu}
\orcid{0000-0001-6184-7641}
\email{hzhual@connect.ust.hk}
\affiliation{
  \institution{Hong Kong University of Science and Technology}
  \country{Hong Kong SAR}
}

\author{Wilfred Siu Hung Ng} 
\orcid{0000-0001-6639-0521}
\email{wng@ust.hk}
\affiliation{
  \institution{Hong Kong University of Science and Technology}
  \country{Hong Kong SAR}
}

\renewcommand{\shortauthors}{Mingshi Xu, Haoren Zhu, and Wilfred Siu Hung Ng}

\usepackage{xcolor}
\usepackage{balance} 

\usepackage[
url=false,
sorting=ynt
]{biblatex}
\usepackage{booktabs}
\usepackage{multirow}
\usepackage{makecell}
\usepackage{float}
\usepackage{enumitem}
\begin{document}

\title{Multi-Item-Query Attention for Stable Sequential Recommendation}

\renewcommand{\shortauthors}{Xu et al.}


\begin{abstract}
The inherent instability and noise in user interaction data challenge sequential recommendation systems. Prevailing masked attention models, relying on a single query from the most recent item, are sensitive to this noise, reducing prediction reliability. We propose the Multi-Item-Query attention mechanism (MIQ-Attn) to enhance model stability and accuracy. MIQ-Attn constructs multiple diverse query vectors from user interactions, effectively mitigating noise and improving consistency. It is designed for easy adoption as a drop-in replacement for existing single-query attention. Experiments show MIQ-Attn significantly improves performance on benchmark datasets.

\end{abstract}



\maketitle

\vspace{-1mm}
\section{INTRODUCTION}
In the dynamic field of recommendation systems, the increasing volume of temporal user-item interaction data presents both new opportunities and significant challenges. As interaction sequences lengthen, they encapsulate more complex behavioral patterns, reflecting both the richness and the variability of user preferences. This trend underscores the critical need for recommendation methodologies that are robust and adaptive to the uncertain and unstable nature of user behaviors.

\begin{figure}[!t]
\setlength{\abovecaptionskip}{3pt}
\includegraphics[width=0.5\textwidth]{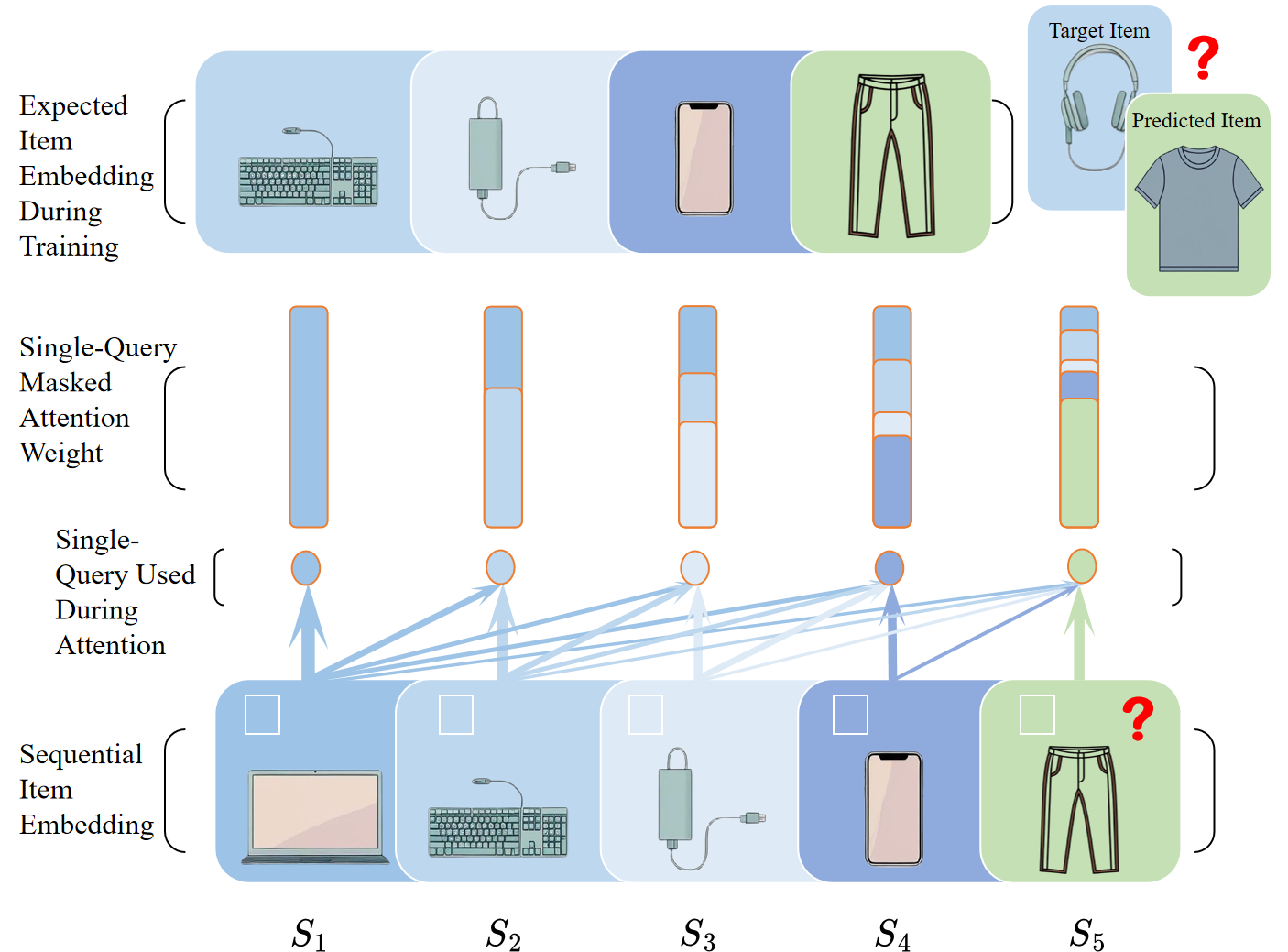}
\caption{
A simplified diagram illustrating the conventional masked attention layer. At each time step, the model computes attention weights using only the query from the most recent item, resulting in a heavily biased output.
}
\label{fig:intro}
\end{figure}

Several approaches have been designed to address the uncertainty in user interaction sequences. Early methods like the Convolutional Sequence Embedding Recommendation Model leveraged larger convolutional kernels to stabilize feature extraction from user interactions \cite{cnnrec_3, cnnrec_2, caser, CNNRec, }. With the attention mechanisms applied in various models \cite{SASRec, attrec, bert4rec,  CL4SRec, S^3Rec, Transformers4Rec, ICLRec, sasbert, }, there was a significant improvement in prediction stability and accuracy due to their ability to consider entire interaction sequences within the input window. Furthermore, some models have integrated probabilistic distributions with the attention mechanism to capture uncertainty or enhance feature representation in item prediction \cite{ PbAttnRNN, DistributionRec, DT4Rec, Degeneration, DCL4Rec, DCRec}. However, these attention mechanisms in sequential recommendation systems rely heavily on the masked attention mechanism to prevent information leakage \cite{Attention}. As a result, despite the attention layer considering various items to aggregate a final prediction, it predominantly generates the query vector based solely on the latest interacted item. which disproportionately amplifies the influence of the most recent item, potentially skewing prediction accuracy. Figure \ref{fig:intro} demonstrates how predictions can become vulnerable when only the most recent item is used for the query.

In response to the challenges of unstable user behaviors in sequential recommendation systems, we introduce a novel Multi-Item-Query attention mechanism (MIQ-Attn) that synthesizes multiple interaction points to construct diversified query item vectors.

While the concept of leveraging multiple queries or modified attention structures bears a superficial resemblance to recent advancements in Large Language Models (LLMs) such as Multi-Query Attention \cite{MQA1029}, Grouped-Query Attention \cite{GQA2023}, or Multi-Head Latent Attention \cite{MLA2024}, our MIQ-Attn is fundamentally different in its motivation, mechanism, and application.
Aforementioned attention mechanisms in LLMs primarily aim to alleviate the substantial computational and memory overhead to enhance training and inference speed.
In contrast, for sequential recommendation tasks, particularly with common benchmark datasets, the computational cost of the attention mechanism itself is often not the primary bottleneck, as discussed in our complexity analysis (Section \ref{sec:complexity_analysis}). 
Instead, the paramount challenge we address is the inherent instability and noise within user interaction sequences by generating multiple, diverse query vectors from recently interacted items, including strategically introduced dummy items.

The MIQ-Attn mitigates the undue influence of any single interaction by distributing attention across a broader context of user interactions. This strategy effectively captures the nuances of user behavior variability, providing more balanced and stable item predictions. This mechanism is meticulously designed to require minimal modifications to existing architectures. It can be seamlessly integrated into existing attention-based models, enhancing their robustness and stability without a substantial system overhaul.

Through rigorous experiments, we compare the performance of the original SASRec \cite{SASRec} with our MIQ-Attn$_{SASRec}$, revealing that our approach substantially improves sequential item predictions. Our findings indicate that MIQ-Attn can significantly improve prediction accuracy for the datasets characterized by long-term complex user behaviors. By leveraging MIQ-Attn in the S$^3$Rec \cite{S^3Rec}, based on our best knowledge,  we achieve state-of-the-art performance on the LastFM \cite{LastFM} dataset. 

The contributions of this paper are stated as follows:
\noindent \begin{itemize}[leftmargin=*]
\item We address the inherent uncertainties within sequential interaction data by diversifying the query generation process.
\item We propose a novel MIQ-Attn mechanism that seamlessly integrates with existing models to enhance prediction accuracy.
\item We empirically validate MIQ-Attn, demonstrating its superiority through improved performance metrics on benchmark datasets.
\end{itemize}

\vspace{-2mm}
\section{METHODOLOGY}

\begin{figure}[t]
\setlength{\abovecaptionskip}{1pt}
\includegraphics[width=0.48\textwidth]{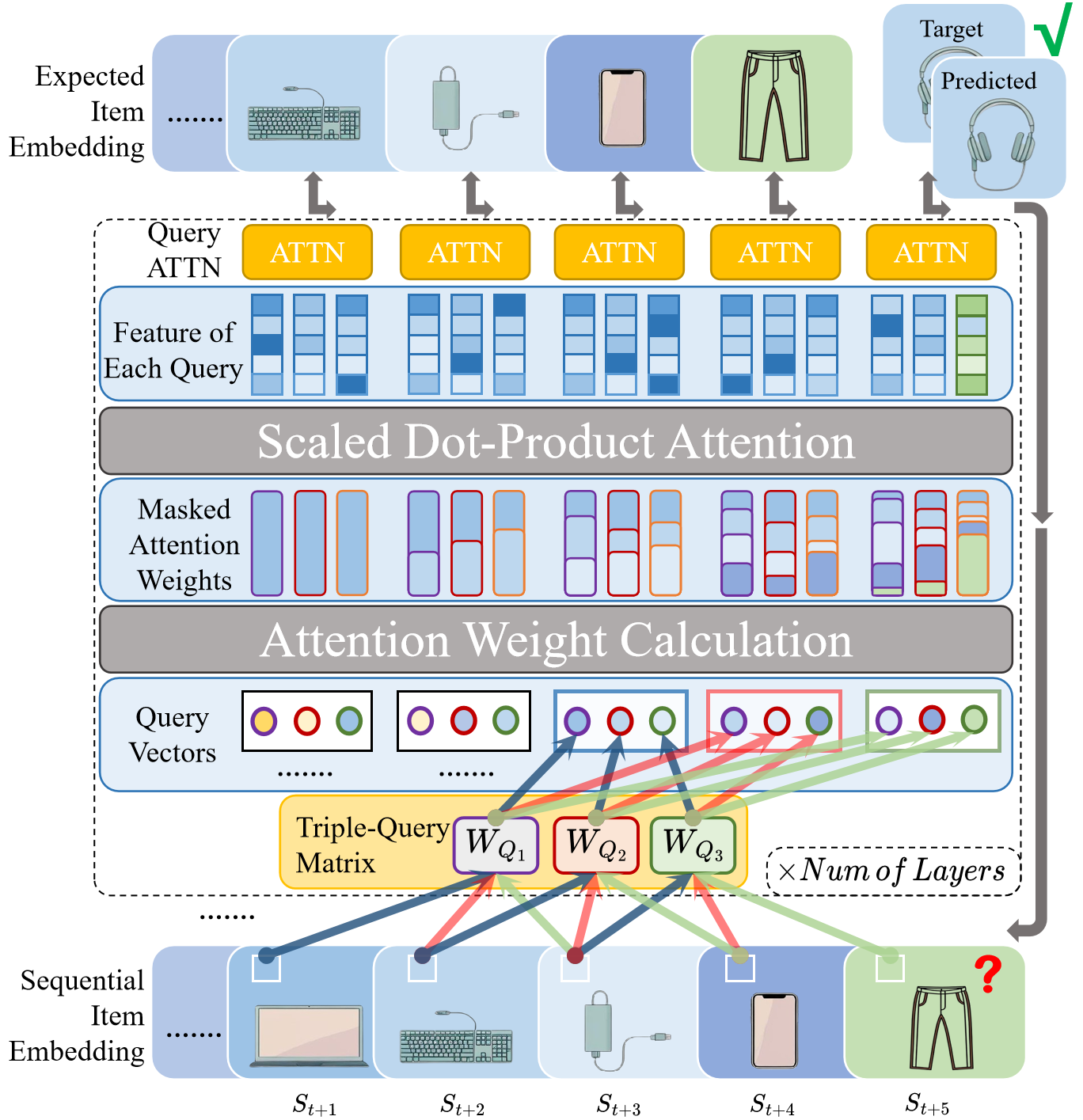}
\vspace{-4mm}
\caption{An overview of the proposed MIQ-Attn framework with query window $m$=3.}\label{fig:overall}
\end{figure}

Sequential recommendation models leveraging attention mechanisms often exhibit sensitivity to the most recent items in a user's interaction history since conventional masked attention mechanisms generate their query vector based solely on the latest interacted item \cite{SASRec}. The MIQ-Attn constructs a more robust and comprehensive prediction by diversifying the query generation process. Figure \ref{fig:overall} depicts the overall architecture of MIQ-Attn. 

\vspace{-2mm}
\subsection{Notation}
Let \( U \) be the set of all users, and \( S \) be the set of all items available for recommendation. Each user \( i \in U \) has an associated sequence of item interactions \( u_i = \{s^i_1, s^i_2, \ldots, s^i_{T_i}\} \), where \( T_i \) represents the length of user \( i \)'s interaction sequence. The primary objective is to predict the next item \( s^i_{T_i+1} \) that user \( i \) is most likely to interact with based on their historical sequence \( u_i \). To manage sequences of varying lengths, all sequences are padded to a uniform length \( T \).

Formally, we define a prediction function, \( f: U \times S^{T} \rightarrow S \), which estimates the next item based on the prior sequence. \( Q \) denotes the query vectors generated within MIQ-Attn, where each \( q_j \in Q \) is a query generated at position \( j \) within the query window $m$.

\vspace{-2mm}

\subsection{Multi-Item-Query Generation}
A key technical challenge in generating multiple queries from an interaction sequence is handling items at different positions consistently, especially initial items that lack sufficient preceding history, and ensuring that each query contributes meaningfully to capturing diverse aspects of user intent. To overcome this, MIQ-Attn employs a two-fold strategy for generating query vectors: utilizing a Query Window Matrix applied to a carefully constructed input sequence that includes dummy items.

\noindent \textbf{Dummy Item Query:}
Dummy items, denoted as \( d_i \), form the set \( D = \{d_1, d_2, \ldots, d_{m-1}\} \), where \( m \) is the size of the query window. These dummy items are introduced at the beginning of sequences to ensure uniformity in the training process across all items in a sequence. By mimicking padding tokens and $[CLS]$ tokens in Bert \cite{Bert} architectures, \( D \) provides initial semantic cues that help stabilize model predictions, especially in sparse data scenarios.
This uniformity is essential as it allows the model to treat the query of the first few items equally to the later ones. 

The expected query generation sequence for user \( i \) at time step \( t \), particularly for the initial items in the sequence, is defined as:
\[
S_{q,t,i} = \left\{
\begin{array}{ll}
\{d_t, d_{t+1}, \ldots, d_{m-1}, s^i_1, \ldots, s^i_t\} & \text{if } t < m \\
\{s^i_{t-m}, \ldots, s^i_t\} & \text{if } m \leq t < T_i
\end{array}
\right.
\]
This structured approach provides a consistent contextual grounding for all items, mitigating the cold-start problem for initial sequence elements and improving prediction stability. Alternative approaches, such as zero-padding or replicating the first item, would not provide learnable semantic cues.

\noindent \textbf{Query Window Matrix:}

Simply using the original embeddings of multiple items from the recent history as queries might not be optimal. Different historical items might signal different aspects of user intent, and a mechanism is needed to transform these source item embeddings into distinct query vectors that can each focus on different facets of the sequence.

Inspired by the multi-head mechanism in Transformers, but applied differently to query generation, we introduce the Query Window Matrix (QWM). The QWM consists of a set of $m$ distinct projection matrices: 
\vspace{-1mm}
\[ W_Q = \{W_{q_1}, W_{q_2}, \ldots, W_{q_m}\} \] 
where each \( W_{q_j} \) in \( \mathbb{R}^{d \times d} \) corresponds to a specific position \( j \) within the query window of size \( m \). This configuration allows for the generation of unique query vectors \( q_j = \hat{E}W_{q_j} \) tailored to each interaction point, ensuring that each query vector captures the nuances of its respective context.

The QWM ensures that while the item embedding matrix \( \hat{E} \) is applied uniformly, the transformation by different matrices \( W_{Q_j} \) at each window position allows for the generation of distinctive and context-sensitive queries. This captures temporal dependencies and also adapts dynamically to positional variations across different interaction segments.

\noindent \textbf{Query Level Attention:}
After generating $m$ distinct query vectors, we obtain $m$ corresponding output feature vectors $O = \{o_1, o_2, \ldots, o_m\}$. Simple averaging or concatenation might not optimally weight the contributions of different queries, as some queries might be more relevant or reliable than others.

To address this, we employ a query-level attention mechanism in which the set of output feature vectors $O$ are treated as inputs to another attention layer.
Let $O_{agg} \in \mathbb{R}^{m \times d}$ be the matrix formed by stacking the $m$ output vectors. We compute aggregated features $F_{agg}$ as:
\vspace{-1.5mm}
\[ \alpha_j = \text{softmax}_j ( (O_{agg} W_K) (o_j W_Q)^T / \sqrt{d_k} ) \] 
\vspace{-3mm}
\[ F_{agg} = \sum_{j=1}^{m} \alpha_j (o_j W_V) \]

where $W_Q, W_K, W_V$ are learnable projection matrices for this query-level attention, and the resulting $F_{agg}$ is the layer output.

\vspace{-2mm}
\subsection{Complexity Analysis}\label{sec:complexity_analysis}

\textbf{Space Complexity:}
In single query attention, the space complexity is \( O(d^2) \). For MIQ-Attn, this increases to \( O(md^2) \), where \( m \) represents the number of queries, and \( d \) is the dimensionality of the embeddings. The additional space required for multiple projection matrices remains minor compared to the total model size, because \( md^2 \) is often substantially smaller than the item embedding matrix with size \( |I|d \), where $|I|$ is the total number of items.

\noindent \textbf{Time Complexity:}
The time complexity for traditional single query attention is \( O(T^2d) \), reflective of the pairwise computation across all sequence elements. With MIQ-Attn, it extends to \( O(mT^2d + m^2Td) \). Although this introduces additional computations, the relatively small size of \( m \) compared to \( T \) and the parallelizable nature of self-attention calculations ensure that the overall training time remains largely unaffected.

In summary, while MIQ-Attn introduces additional complexity in terms of space and time, the increases are manageable and justified by the enhanced modeling capabilities and adaptability, especially given the efficiencies of parallel computation \cite{Parallelization}.

\vspace{-1mm}
\section{EXPERIMENTATION}
This section addresses three research issues: \textbf{R1:} Can MIQ-Attn improve model performance on different datasets? \textbf{R2:} How to select an appropriate hyperparameter query window $m$, given the distinct properties of the datasets? \textbf{R3:} Since many models utilize attention and their architecture varies, can MIQ-Attn enhance the performance of more advanced attention-based models?

\vspace{-2mm}
\subsection{Experimental Setup}

\begin{table}[!t]
\small
\centering
\setlength{\abovecaptionskip}{1pt}
\caption{Dataset statistics (after preprocessing)}
\label{tab:dataset_stats}
\begin{tabular}{lccccc}
\toprule
\textbf{Dataset} & \textbf{\#users} & \textbf{\#items} & \thead{\textbf{avg.} \\ \textbf{actions} \\ \textbf{/user}} & \thead{\textbf{avg.} \\ \textbf{actions} \\ \textbf{/item}} & \textbf{\#actions} \\
\midrule
Amazon Beauty & 52,024 & 57,289 & 7.6 & 6.9 & 0.4M \\
MovieLens-1M & 6,040 & 3,416 & 163.5 & 289.1 & 1.0M \\
LastFM & 1,090 & 3,646 & 48.2 & 14.4 & 52,551 \\
\bottomrule
\end{tabular}
\vspace{-4mm}
\end{table}

\begin{table*}[ht]
\small
\centering
\setlength{\abovecaptionskip}{1pt}
\caption{Comparative performance on the Beauty and ML-1M datasets. The subscript $m$ of MIQ-Attn$_m$ represents the size of the query window used. The best and second-best methods in each row are shown in bold and underlined, respectively.}
\label{Beauty&ML}
\begin{tabular}{@{}clcccccccccc@{}}
\toprule
\multirow{2}{*}{@n} & \multirow{2}{*}{Metric} & \multicolumn{5}{c}{Beauty} & \multicolumn{5}{c}{ML-1M}\\
\cmidrule(lr){3-7}\cmidrule(lr){8-12}
 & & SASRec & MIQ-Attn$_{2}$ & MIQ-Attn$_{3}$ & MIQ-Attn$_{5}$ & MIQ-Attn$_{10}$ &  SASRec & MIQ-Attn$_{3}$ & MIQ-Attn$_{5}$ & MIQ-Attn$_{10}$ & MIQ-Attn$_{15}$ \\ 
\midrule

 \multirow{2}{*}{5}
 & HIT   & 0.0165 & \underline{0.0175} & 0.0123 & \textbf{0.0195} & 0.0021 & 0.0850 & 0.0910 & 0.0880 & \textbf{0.1030} & \underline{0.0960} \\
 & NDCG  & 0.0093 & \underline{0.0096} & 0.0066 & \textbf{0.0104} & 0.0010 & \underline{0.0508} & 0.0451 & 0.0453 & \textbf{0.0515} & 0.0471 \\
 \multirow{2}{*}{10}
 & HIT   & \textbf{0.0267} & \underline{0.0258} & 0.0246 & 0.0246 & 0.0021 & 0.1570 & 0.1610 & \underline{0.1780} & \textbf{0.1790} & 0.1660 \\
 & NDCG  & \textbf{0.0125} & \underline{0.0123} & 0.0104 & 0.0121 & 0.0010 & 0.0739 & 0.0677 & \underline{0.0742} & \textbf{0.0759} & 0.0692 \\
 \multirow{2}{*}{20}
 & HIT   & 0.0412 & 0.0402 & \textbf{0.0494} & \underline{0.0421} & 0.0052 & 0.2610 & 0.2770 & \underline{0.2740} & \textbf{0.2910} & 0.2590 \\
 & NDCG  & 0.0162 & 0.0160 & \textbf{0.0167} & \underline{0.0165} & 0.0018 & \underline{0.0999} & 0.0967 & 0.0984 & \textbf{0.1041} & 0.0926 \\
\bottomrule
\end{tabular}
\vspace{-2mm}
\end{table*}

\begin{table*}[ht]
\small
\centering
\setlength{\abovecaptionskip}{1pt}
\caption{Performance on the LastFM Dataset Across Different Models. The best and second-best methods in each row are shown in bold and underlined, respectively. Query window $m = 5$ in this table.  \% Imp1 refers to improvement from MIQ-Attn$_{SASRec}$ vs. SASRec, \% Imp2 refers to improvement from second best to best. $^*$ means the improvement is statistically significant. }
\label{Last_FM_SOTA}
\begin{tabular}{@{}clcccccccccc@{}}
\toprule
\multirow{2}{*}{@n} & \multirow{2}{*}{Metric} & \multicolumn{5}{c}{Type 1: No Pretrain, Item Sequence Only} &  & \multicolumn{4}{c}{Type 2: Pretrain, Other Features Included} \\
\cmidrule(lr){3-7} \cmidrule(lr){9-12}
 & & GRU4Rec$_F$ & SRGNN$_F$ & SASRec$_F$ & MIQ-Attn$_{SASRec_F}$ & Imp1 & & S$^3$Rec & S$^3$Rec$_F$ & MIQ-Attn$_{S^3Rec_F}$ & Imp2\\ 
\midrule

 \multirow{2}{*}{5}
 & HIT  & 0.0339 & 0.0303 & 0.0303 & 0.0376 & 24.09\%$^*$ & & 0.0413 & \underline{0.0431} & \textbf{0.0541} & 25.52\%$^*$ \\
 & NDCG & 0.0233 & 0.0200 & 0.0161 & 0.0194 & 20.50\%$^*$ & & \underline{0.0264} & 0.0250 & \textbf{0.0365} & 38.26\%$^*$ \\
 \multirow{2}{*}{10}
 & HIT  & 0.0468 & 0.0468 & 0.0486 & 0.0578 & 18.93\%$^*$ & & \underline{0.0752} & 0.0743 & \textbf{0.0844} & 12.23\%$^*$ \\
 & NDCG & 0.0274 & 0.0253 & 0.0221 & 0.0261 & 18.10\%$^*$ & & \underline{0.0371} & 0.0349 & \textbf{0.0459} & 23.72\%$^*$ \\
 \multirow{2}{*}{20}
 & HIT  & 0.0624 & 0.0596 & 0.0826 & 0.0881 & 6.66\%$^*$  & & 0.1193 & \underline{0.1266} & \textbf{0.1294} & 2.21\% \\
 & NDCG & 0.0313 & 0.0285 & 0.0305 & 0.0337 & 10.49\%$^*$ & & 0.0480 & \underline{0.0481} & \textbf{0.0572} & 18.92\%$^*$ \\
\bottomrule
\end{tabular}
\vspace{-3mm}
\end{table*}

\noindent \textbf{Datasets:} We used three benchmark datasets in our experiments: \textbf{Beauty} \cite{Beauty}, \textbf{ML-1M} \cite{ml-1m}, and \textbf{LastFM} \cite{LastFM}. Due to space limitation, the Beauty and ML-1M were used to evaluate the effectiveness of MIQ-Attn mechanism across diverse domains and conduct ablation study. 
On the LastFM dataset, we applied the MIQ-Attn mechanism to two attention models to verify its capacity on more advanced models. For Beauty and ML-1M, we adhered to the original configurations from SASRec, whereas for LastFM, we followed the settings from S$^3$Rec. Dataset statistics are detailed in Table \ref{tab:dataset_stats}.

\noindent \textbf{Evaluation Metrics:} We employ two widely recognized metrics, the top-k Hit Ratio (\textbf{HR@k}) and top-k Normalized Discounted Cumulative Gain (\textbf{NDCG@k}). We report results for HR and NDCG at cutoffs of 5, 10, and 20. Consistent with best practices in the field \cite{SASRec, S^3Rec, improvingRec}, we adopt a leave-one-out evaluation strategy.  Specifically, for each user interaction sequence, the last item serves as the test sample, the penultimate item as the validation sample, and all preceding items as training data. To ensure robust comparisons and mitigate the variability due to traditional sampling methods, we employ the full-corpus method for negative sample selection, which uses all items that did not interact with the user as the negative samples, and is widely accepted in the latest works \cite{reviewRec2022,  EEss, improvingRec, hstu,}. 

\noindent \textbf{Parameter Settings:} For experimental settings, all the benchmarks use their original hyperparameters which show the best performance. For MIQ-Attn\footnote{https://github.com/mxuax/MQ-Attn2025}, we adjust the item embedding size and the dropout rate according to the query window size. Our model utilizes the PyTorch library on a GeForce RTX 4080 GPU.

\vspace{-2mm}
\subsection{Experimental Results}
Table \ref{Beauty&ML} illustrates the comparative efficacy of the original SASRec model with the MIQ-Attn version (\textbf{R1}). When applied to the ML-1M dataset, our enhanced model demonstrates a notable improvement in all performance metrics. However, the results of the Beauty dataset were less pronounced. This discrepancy could be attributed to the shorter average interaction length per user in the Beauty dataset. The introduction of additional dummy item sequences may have disproportionately influenced the original sequence, leading to suboptimal outcomes. This effect was obvious when query window $m = 10$, which exceeds the average sequence length, resulting in a drastically drop in model performance. 

Concerning \textbf{R2}, our investigation into the optimal size for the query window $m$ suggests that it should approximately equate to 1/10 of the average actions per user as shown in Table \ref{Beauty&ML} \textbf{ML-1M}, this is also a clue since for \textbf{Beauty}, the average length is around 8, thus the optimal query window $m$ should be around 1, which means the unsuitability of using MIQ-Attn.

Table \ref{Last_FM_SOTA} presents the results of applying MIQ-Attn to two model variations: SASRec$_F$ and S$^3$Rec$_F$. In this context, the subscript "$_F$" denotes a variant of the foundational model that integrates a novel loss function with a cumulative temporal term \cite{improvingRec}. Additionally, we include results from the original S$^3$Rec\cite{S^3Rec} implementation, GRU4Rec$_F$\cite{GRU4Rec}, and SRGNN$_F$\cite{SRGNN} to ensure robust assessment. 
Notably, MIQ-Attn achieves the highest performance across both \textbf{Type 1:} the basic model, which utilizes only item sequence data without pretraining, and \textbf{Type 2:} the more sophisticated models that employ a self-supervised approach to leverage intrinsic data correlations as additional signals through pretrain models, subsequently fine-tuned on sequential data \cite{S^3Rec}. It is noteworthy that MIQ-Attn${S^3Rec_F}$ has achieved state-of-the-art performance on the LastFM dataset. This comprehensive evaluation underscores the adaptability and performance enhancement offered by MIQ-Attn (\textbf{R3}). 

\begin{figure}[!t]
\setlength{\abovecaptionskip}{1pt}
\includegraphics[width=0.48\textwidth]{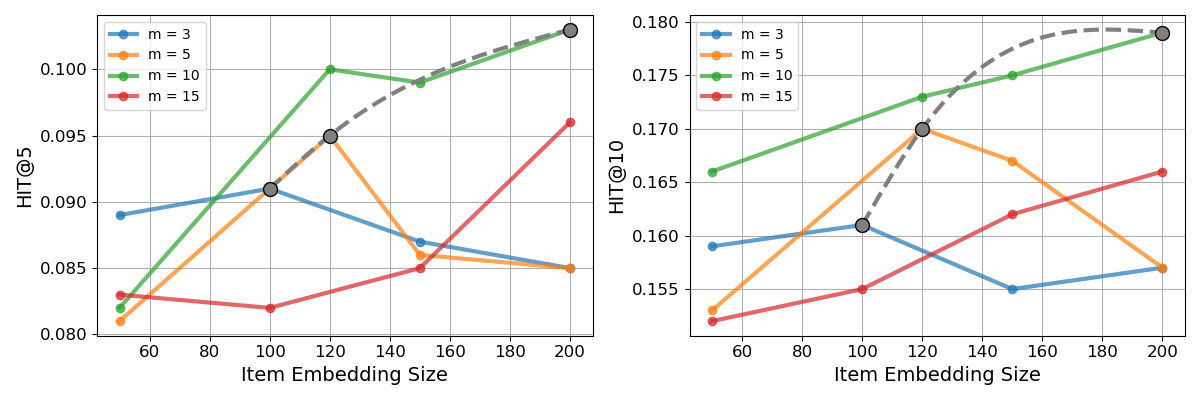}
\caption{The relationships between model performance and item embedding size with different query window $m$ are shown. Each point in the figure corresponds to the highest achieved metric score under specific hyperparameters.} \label{fig:hyperparameter}
\vspace{-5mm}
\end{figure}
\vspace{-3mm}
\subsection{Hyperparameter Analysis}

In exploring the optimal settings for MIQ-Attn, we analyzed the interplay between the query window size
$m$ and the item embedding size. Due to space constraints, we evaluate on the ML-1M dataset with the HIT@5 and HIT@10 metrics.  Figure \ref{fig:hyperparameter} shows the results, the grey dots show the best embedding size given $m$ and the dashed spline connecting them indicates that optimal item embedding sizes increase in tandem with the query window, particularly when the query window size is within one-tenth of the average interaction length. This suggests a scalable relationship between the query window and embedding dimensions that merits further exploration for parameter-tuning to enhance performance in recommendation systems. The propensity for larger embedding sizes to improve model performance may be attributed to the enhanced representation capabilities afforded by MIQ-Attn.

\vspace{-2mm}
\section{CONCLUSION}

In this study, we introduce a novel Multi-Item-Query attention mechanism for sequential recommendation systems and validated its effectiveness through systematic experimentation. Our empirical findings demonstrate that MIQ-Attn significantly improves the accuracy of recommendations by addressing a critical limitation in traditional masked attention mechanisms.
Future work focuses on two main aspects: (1) exploring a more dynamic query mechanism that can adaptively select the optimal query window based on the real-time interaction sequence of users, and (2) delving deeper into the feature vectors generated from different queries to investigate their potential in reflecting the uncertainty properties of users.

\balance 
\printbibliography

@misc{MLA2024,
      title={DeepSeek-V2: A Strong, Economical, and Efficient Mixture-of-Experts Language Model}, 
      author={DeepSeek-AI and Aixin Liu and Bei Feng and Bin Wang and Bingxuan Wang and Bo Liu and Chenggang Zhao and Chengqi Dengr and Chong Ruan and Damai Dai and Daya Guo and Dejian Yang and Deli Chen and Dongjie Ji and Erhang Li and Fangyun Lin and Fuli Luo and Guangbo Hao and Guanting Chen and Guowei Li and H. Zhang and Hanwei Xu and Hao Yang and Haowei Zhang and Honghui Ding and Huajian Xin and Huazuo Gao and Hui Li and Hui Qu and J. L. Cai and Jian Liang and Jianzhong Guo and Jiaqi Ni and Jiashi Li and Jin Chen and Jingyang Yuan and Junjie Qiu and Junxiao Song and Kai Dong and Kaige Gao and Kang Guan and Lean Wang and Lecong Zhang and Lei Xu and Leyi Xia and Liang Zhao and Liyue Zhang and Meng Li and Miaojun Wang and Mingchuan Zhang and Minghua Zhang and Minghui Tang and Mingming Li and Ning Tian and Panpan Huang and Peiyi Wang and Peng Zhang and Qihao Zhu and Qinyu Chen and Qiushi Du and R. J. Chen and R. L. Jin and Ruiqi Ge and Ruizhe Pan and Runxin Xu and Ruyi Chen and S. S. Li and Shanghao Lu and Shangyan Zhou and Shanhuang Chen and Shaoqing Wu and Shengfeng Ye and Shirong Ma and Shiyu Wang and Shuang Zhou and Shuiping Yu and Shunfeng Zhou and Size Zheng and T. Wang and Tian Pei and Tian Yuan and Tianyu Sun and W. L. Xiao and Wangding Zeng and Wei An and Wen Liu and Wenfeng Liang and Wenjun Gao and Wentao Zhang and X. Q. Li and Xiangyue Jin and Xianzu Wang and Xiao Bi and Xiaodong Liu and Xiaohan Wang and Xiaojin Shen and Xiaokang Chen and Xiaosha Chen and Xiaotao Nie and Xiaowen Sun and Xiaoxiang Wang and Xin Liu and Xin Xie and Xingkai Yu and Xinnan Song and Xinyi Zhou and Xinyu Yang and Xuan Lu and Xuecheng Su and Y. Wu and Y. K. Li and Y. X. Wei and Y. X. Zhu and Yanhong Xu and Yanping Huang and Yao Li and Yao Zhao and Yaofeng Sun and Yaohui Li and Yaohui Wang and Yi Zheng and Yichao Zhang and Yiliang Xiong and Yilong Zhao and Ying He and Ying Tang and Yishi Piao and Yixin Dong and Yixuan Tan and Yiyuan Liu and Yongji Wang and Yongqiang Guo and Yuchen Zhu and Yuduan Wang and Yuheng Zou and Yukun Zha and Yunxian Ma and Yuting Yan and Yuxiang You and Yuxuan Liu and Z. Z. Ren and Zehui Ren and Zhangli Sha and Zhe Fu and Zhen Huang and Zhen Zhang and Zhenda Xie and Zhewen Hao and Zhihong Shao and Zhiniu Wen and Zhipeng Xu and Zhongyu Zhang and Zhuoshu Li and Zihan Wang and Zihui Gu and Zilin Li and Ziwei Xie},
      year={2024},
      eprint={2405.04434},
      archivePrefix={arXiv},
      primaryClass={cs.CL},
      url={https://arxiv.org/abs/2405.04434}, 
}

@misc{GQA2023,
      title={GQA: Training Generalized Multi-Query Transformer Models from Multi-Head Checkpoints}, 
      author={Joshua Ainslie and James Lee-Thorp and Michiel de Jong and Yury Zemlyanskiy and Federico Lebrón and Sumit Sanghai},
      year={2023},
      eprint={2305.13245},
      archivePrefix={arXiv},
      primaryClass={cs.CL},
      url={https://arxiv.org/abs/2305.13245}, 
}

@article{MQA1029,
  author       = {Noam Shazeer},
  title        = {Fast Transformer Decoding: One Write-Head is All You Need},
  journal      = {CoRR},
  volume       = {abs/1911.02150},
  year         = {2019},
  url          = {http://arxiv.org/abs/1911.02150},
  eprinttype    = {arXiv},
  eprint       = {1911.02150},
  bibsource    = {dblp computer science bibliography, https://dblp.org}
}

@INPROCEEDINGS{LastFM,
  author = {Thierry Bertin-Mahieux and Daniel P.W. Ellis and Brian Whitman and Paul Lamere},
  title = {The Million Song Dataset},
  booktitle = {{Proceedings of the 12th International Conference on Music Information
	Retrieval ({ISMIR} 2011)}},
  year = {2011}
}

@inproceedings{sasbert, series={RecSys ’23},
   title={Turning Dross Into Gold Loss: is BERT4Rec really better than SASRec?},
   url={http://dx.doi.org/10.1145/3604915.3610644},
   DOI={10.1145/3604915.3610644},
   booktitle={Proceedings of the 17th ACM Conference on Recommender Systems},
   publisher={ACM},
   author={Klenitskiy, Anton and Vasilev, Alexey},
   year={2023},
   month=sep, pages={1120–1125},
   collection={RecSys ’23} }

@inproceedings{EEss,
author = {Petrov, Aleksandr and Macdonald, Craig},
title = {Effective and Efficient Training for Sequential Recommendation using Recency Sampling},
year = {2022},
isbn = {9781450392785},
publisher = {Association for Computing Machinery},
address = {New York, NY, USA},
url = {https://doi.org/10.1145/3523227.3546785},
doi = {10.1145/3523227.3546785},
booktitle = {Proceedings of the 16th ACM Conference on Recommender Systems},
pages = {81–91},
numpages = {11},
location = {Seattle, WA, USA},
series = {RecSys '22}
}

@article{SRGNN,
  author       = {Shu Wu and
                  Yuyuan Tang and
                  Yanqiao Zhu and
                  Liang Wang and
                  Xing Xie and
                  Tieniu Tan},
  title        = {Session-based Recommendation with Graph Neural Networks},
  journal      = {CoRR},
  volume       = {abs/1811.00855},
  year         = {2018},
  url          = {http://arxiv.org/abs/1811.00855},
  eprinttype    = {arXiv},
  eprint       = {1811.00855},
  bibsource    = {dblp computer science bibliography, https://dblp.org}
}

@misc{GRU4Rec,
      title={Session-based Recommendations with Recurrent Neural Networks}, 
      author={Balázs Hidasi and Alexandros Karatzoglou and Linas Baltrunas and Domonkos Tikk},
      year={2016},
      eprint={1511.06939},
      archivePrefix={arXiv},
      primaryClass={cs.LG},
      url={https://arxiv.org/abs/1511.06939}, 
}

@misc{reviewRec2022,
      title={A Systematic Review and Replicability Study of BERT4Rec for Sequential Recommendation}, 
      author={Aleksandr Petrov and Craig Macdonald},
      year={2022},
      eprint={2207.07483},
      archivePrefix={arXiv},
      primaryClass={cs.IR},
      url={https://arxiv.org/abs/2207.07483}, 
}

@misc{hstu,
      title={Actions Speak Louder than Words: Trillion-Parameter Sequential Transducers for Generative Recommendations}, 
      author={Jiaqi Zhai and Lucy Liao and Xing Liu and Yueming Wang and Rui Li and Xuan Cao and Leon Gao and Zhaojie Gong and Fangda Gu and Michael He and Yinghai Lu and Yu Shi},
      year={2024},
      eprint={2402.17152},
      archivePrefix={arXiv},
      primaryClass={cs.LG},
      url={https://arxiv.org/abs/2402.17152}, 
}

@misc{improvingRec,
      title={Improving Sequential Recommendation Models with an Enhanced Loss Function},
      author={Fangyu Li and Shenbao Yu and Feng Zeng and Fang Yang},
      year={2023},
      eprint={2301.00979},
      archivePrefix={arXiv},
      primaryClass={cs.IR}
}

@article{ml-1m,
  title={The movielens datasets: History and context},
  author={Harper, F Maxwell and Konstan, Joseph A},
  journal={Acm transactions on interactive intelligent systems (tiis)},
  volume={5},
  number={4},
  pages={1--19},
  year={2015},
  publisher={Acm New York, NY, USA}
}

@article{Beauty,
  author       = {Julian J. McAuley and
                  Christopher Targett and
                  Qinfeng Shi and
                  Anton van den Hengel},
  title        = {Image-based Recommendations on Styles and Substitutes},
  journal      = {CoRR},
  volume       = {abs/1506.04757},
  year         = {2015},
  url          = {http://arxiv.org/abs/1506.04757},
  eprinttype    = {arXiv},
  eprint       = {1506.04757},
  bibsource    = {dblp computer science bibliography, https://dblp.org}
}

@article{Attention,
  author       = {Ashish Vaswani and
                  Noam Shazeer and
                  Niki Parmar and
                  Jakob Uszkoreit and
                  Llion Jones and
                  Aidan N. Gomez and
                  Lukasz Kaiser and
                  Illia Polosukhin},
  title        = {Attention Is All You Need},
  journal      = {CoRR},
  volume       = {abs/1706.03762},
  year         = {2017},
  url          = {http://arxiv.org/abs/1706.03762},
  eprinttype    = {arXiv},
  eprint       = {1706.03762},
  bibsource    = {dblp computer science bibliography, https://dblp.org}
}

@inproceedings{DT4Rec,
  title = {Modeling {{Sequences}} as {{Distributions}} with {{Uncertainty}} for {{Sequential Recommendation}}},
  booktitle = {Proceedings of the 30th {{ACM International Conference}} on {{Information}} \& {{Knowledge Management}}},
  author = {Fan, Ziwei and Liu, Zhiwei and Wang, Shen and Zheng, Lei and Yu, Philip S.},
  year = {2021},
  month = oct,
  pages = {3019--3023},
  publisher = {ACM},
  address = {Virtual Event Queensland Australia},
  doi = {10.1145/3459637.3482145},
  urldate = {2024-07-11},
  isbn = {978-1-4503-8446-9},
  langid = {english}
}

@inproceedings{Transformers4Rec,
author = {de Souza Pereira Moreira, Gabriel and Rabhi, Sara and Lee, Jeong Min and Ak, Ronay and Oldridge, Even},
title = {Transformers4Rec: Bridging the Gap between NLP and Sequential / Session-Based Recommendation},
year = {2021},
isbn = {9781450384582},
publisher = {Association for Computing Machinery},
address = {New York, NY, USA},
url = {https://doi.org/10.1145/3460231.3474255},
doi = {10.1145/3460231.3474255},
booktitle = {Proceedings of the 15th ACM Conference on Recommender Systems},
pages = {143–153},
numpages = {11},
location = {Amsterdam, Netherlands},
series = {RecSys '21}
}

@inproceedings{cnnrec_3,
author = {Kim, Donghyun and Park, Chanyoung and Oh, Jinoh and Lee, Sungyoung and Yu, Hwanjo},
title = {Convolutional Matrix Factorization for Document Context-Aware Recommendation},
year = {2016},
isbn = {9781450340359},
publisher = {Association for Computing Machinery},
address = {New York, NY, USA},
url = {https://doi.org/10.1145/2959100.2959165},
doi = {10.1145/2959100.2959165},
booktitle = {Proceedings of the 10th ACM Conference on Recommender Systems},
pages = {233–240},
numpages = {8},
location = {Boston, Massachusetts, USA},
series = {RecSys '16}
}

@inproceedings{cnnrec_2,
author = {Zheng, Lei and Noroozi, Vahid and Yu, Philip S.},
title = {Joint Deep Modeling of Users and Items Using Reviews for Recommendation},
year = {2017},
isbn = {9781450346757},
publisher = {Association for Computing Machinery},
address = {New York, NY, USA},
url = {https://doi.org/10.1145/3018661.3018665},
doi = {10.1145/3018661.3018665},
booktitle = {Proceedings of the Tenth ACM International Conference on Web Search and Data Mining},
pages = {425–434},
numpages = {10},
location = {Cambridge, United Kingdom},
series = {WSDM '17}
}

@article{CL4SRec,
  author       = {Xu Xie and
                  Fei Sun and
                  Zhaoyang Liu and
                  Jinyang Gao and
                  Bolin Ding and
                  Bin Cui},
  title        = {Contrastive Pre-training for Sequential Recommendation},
  journal      = {CoRR},
  volume       = {abs/2010.14395},
  year         = {2020},
  url          = {https://arxiv.org/abs/2010.14395},
  eprinttype    = {arXiv},
  eprint       = {2010.14395},
  bibsource    = {dblp computer science bibliography, https://dblp.org}
}

@misc{DCL4Rec,
  title = {Dcl4rec: {{An Effective Debiased Contrastive Learning Framework}} for {{Long-Tail Sequential Recommendation}}},
  shorttitle = {Dcl4rec},
  author = {Hong, Yingfei and Yuan, Xiaoqun and Li, Xuhui},
  year = {2023},
  doi = {10.2139/ssrn.4558746},
  urldate = {2024-09-24},
  langid = {english}
}

@misc{SASRec,
  title = {Self-{{Attentive Sequential Recommendation}}},
  author = {Kang, Wang-Cheng and McAuley, Julian},
  year = {2018},
  month = aug,
  number = {arXiv:1808.09781},
  eprint = {1808.09781},
  primaryclass = {cs},
  publisher = {arXiv},
  urldate = {2024-09-24},
  archiveprefix = {arXiv},
  langid = {english}
}

@inproceedings{Parallelization,
author = {Narayanan, Deepak and Shoeybi, Mohammad and Casper, Jared and LeGresley, Patrick and Patwary, Mostofa and Korthikanti, Vijay and Vainbrand, Dmitri and Kashinkunti, Prethvi and Bernauer, Julie and Catanzaro, Bryan and Phanishayee, Amar and Zaharia, Matei},
title = {Efficient large-scale language model training on GPU clusters using megatron-LM},
year = {2021},
isbn = {9781450384421},
publisher = {Association for Computing Machinery},
address = {New York, NY, USA},
url = {https://doi.org/10.1145/3458817.3476209},
doi = {10.1145/3458817.3476209},
booktitle = {Proceedings of the International Conference for High Performance Computing, Networking, Storage and Analysis},
articleno = {58},
numpages = {15},
location = {St. Louis, Missouri},
series = {SC '21}
}

@article{Bert,
  author       = {Jacob Devlin and
                  Ming{-}Wei Chang and
                  Kenton Lee and
                  Kristina Toutanova},
  title        = {{BERT:} Pre-training of Deep Bidirectional Transformers for Language
                  Understanding},
  journal      = {CoRR},
  volume       = {abs/1810.04805},
  year         = {2018},
  url          = {http://arxiv.org/abs/1810.04805},
  eprinttype    = {arXiv},
  eprint       = {1810.04805},
  bibsource    = {dblp computer science bibliography, https://dblp.org}
}

@article{CNNRec,
author = {Yang, Dan and Zhang, Jing and Wang, Sifeng and Zhang, XueDong},
title = {A Time-Aware CNN-Based Personalized Recommender System},
journal = {Complexity},
volume = {2019},
number = {1},
pages = {9476981},
doi = {https://doi.org/10.1155/2019/9476981},
url = {https://onlinelibrary.wiley.com/doi/abs/10.1155/2019/9476981},
eprint = {https://onlinelibrary.wiley.com/doi/pdf/10.1155/2019/9476981},
year = {2019}
}

@article{ICLRec,
  title={Intent Contrastive Learning for Sequential Recommendation},
  author={Chen, Yongjun and Liu, Zhiwei and Li, Jia and McAuley, Julian and Xiong, Caiming},
  journal={arXiv preprint arXiv:2202.02519},
  year={2022}
}

@inproceedings{DCRec,
  title = {Debiased {{Contrastive Learning}} for {{Sequential Recommendation}}},
  booktitle = {Proceedings of the {{ACM Web Conference}} 2023},
  author = {Yang, Yuhao and Huang, Chao and Xia, Lianghao and Huang, Chunzhen and Luo, Da and Lin, Kangyi},
  year = {2023},
  month = apr,
  eprint = {2303.11780},
  primaryclass = {cs},
  pages = {1063--1073},
  doi = {10.1145/3543507.3583361},
  urldate = {2024-10-31},
  archiveprefix = {arXiv},
  langid = {english}
}

@misc{caser,
  title = {Personalized {{Top-N Sequential Recommendation}} via {{Convolutional Sequence Embedding}}},
  author = {Tang, Jiaxi and Wang, Ke},
  year = {2018},
  month = sep,
  number = {arXiv:1809.07426},
  eprint = {1809.07426},
  primaryclass = {cs},
  publisher = {arXiv},
  doi = {10.48550/arXiv.1809.07426},
  urldate = {2024-11-27},
  archiveprefix = {arXiv},
  langid = {english}
}

@inproceedings{Degeneration,
  title = {Contrastive {{Learning}} for {{Representation Degeneration Problem}} in {{Sequential Recommendation}}},
  booktitle = {Proceedings of the {{Fifteenth ACM International Conference}} on {{Web Search}} and {{Data Mining}}},
  author = {Qiu, Ruihong and Huang, Zi and Yin, Hongzhi and Wang, Zijian},
  year = {2022},
  month = feb,
  eprint = {2110.05730},
  primaryclass = {cs},
  pages = {813--823},
  doi = {10.1145/3488560.3498433},
  urldate = {2024-11-06},
  archiveprefix = {arXiv},
  langid = {english}
}

@inproceedings{DistributionRec,
author = {Zheng, Lei and Li, Chaozhuo and Lu, Chun-Ta and Zhang, Jiawei and Yu, Philip S.},
title = {Deep Distribution Network: Addressing the Data Sparsity Issue for Top-N Recommendation},
year = {2019},
isbn = {9781450361729},
publisher = {Association for Computing Machinery},
address = {New York, NY, USA},
url = {https://doi.org/10.1145/3331184.3331330},
doi = {10.1145/3331184.3331330},
booktitle = {Proceedings of the 42nd International ACM SIGIR Conference on Research and Development in Information Retrieval},
pages = {1081–1084},
numpages = {4},
location = {Paris, France},
series = {SIGIR'19}
}

@article{PbAttnRNN,
    doi = {10.1371/journal.pone.0223967},
    author = {Zhang, Weiwei AND Liu, Fangai AND Xu, Daomeng AND Jiang, Lu},
    journal = {PLOS ONE},
    publisher = {Public Library of Science},
    title = {Recommendation system in social networks with topical attention and probabilistic matrix factorization},
    year = {2019},
    month = {10},
    volume = {14},
    url = {https://doi.org/10.1371/journal.pone.0223967},
    pages = {1-18},
    number = {10},

}

@article{attrec,
  author       = {Shuai Zhang and
                  Yi Tay and
                  Lina Yao and
                  Aixin Sun},
  title        = {Next Item Recommendation with Self-Attention},
  journal      = {CoRR},
  volume       = {abs/1808.06414},
  year         = {2018},
  url          = {http://arxiv.org/abs/1808.06414},
  eprinttype    = {arXiv},
  eprint       = {1808.06414},
  bibsource    = {dblp computer science bibliography, https://dblp.org}
}

@article{bert4rec,
  author       = {Fei Sun and
                  Jun Liu and
                  Jian Wu and
                  Changhua Pei and
                  Xiao Lin and
                  Wenwu Ou and
                  Peng Jiang},
  title        = {BERT4Rec: Sequential Recommendation with Bidirectional Encoder Representations
                  from Transformer},
  journal      = {CoRR},
  volume       = {abs/1904.06690},
  year         = {2019},
  url          = {http://arxiv.org/abs/1904.06690},
  eprinttype    = {arXiv},
  eprint       = {1904.06690},
  bibsource    = {dblp computer science bibliography, https://dblp.org}
}

@article{S^3Rec,
  author       = {Kun Zhou and
                  Hui Wang and
                  Wayne Xin Zhao and
                  Yutao Zhu and
                  Sirui Wang and
                  Fuzheng Zhang and
                  Zhongyuan Wang and
                  Ji{-}Rong Wen},
  title        = {S3-Rec: Self-Supervised Learning for Sequential Recommendation
                  with Mutual Information Maximization},
  journal      = {CoRR},
  volume       = {abs/2008.07873},
  year         = {2020},
  url          = {https://arxiv.org/abs/2008.07873},
  eprinttype    = {arXiv},
  eprint       = {2008.07873},
  bibsource    = {dblp computer science bibliography, https://dblp.org}
}

@String{Computing = "Computing" }

@String{Computer = "{IEEE} Computer" }

@ArtifactSoftware{R,
    title = {R: A Language and Environment for Statistical Computing},
    author = {{R Core Team}},
    organization = {R Foundation for Statistical Computing},
    address = {Vienna, Austria},
    year = {2019},
    url = {https://www.R-project.org/},
}

\end{document}